# The Comparative Exploration of the Ice Giant Planets with Twin Spacecraft: Unveiling the History of our Solar System

Diego Turrini[1*], Romolo Politi[1], Roberto Peron[1], Davide Grassi[1], Christina Plainaki[1], Mauro Barbieri[2], David M. Lucchesi[1], Gianfranco Magni[1], Francesca Altieri[1], Valeria Cottini[3], Nicolas Gorius[4], Patrick Gaulme[5,6], François-Xavier Schmider[7], Alberto Adriani[1], Giuseppe Piccioni[1]

1. Institute for Space Astrophysics and Planetology INAF-IAPS, Italy.
2. Center of Studies and Activities for Space CISAS, University of Padova, Italy.
3. University of Maryland, USA.
4. Catholic University of America, USA
5. Department of Astronomy, New Mexico State University, P.O. Box 30001, MSC 4500, Las Cruces, NM 88003-8001, USA
6. Apache Point Observatory, 2001 Apache Point Road, P.O. Box 59, Sunspot, NM 88349, USA
7. Laboratoire Lagrange, Observatoire de la Côte d'Azur, France

## Abstract

In the course of the selection of the scientific themes for the second and third L-class missions of the Cosmic Vision 2015-2025 program of the European Space Agency, the exploration of the ice giant planets Uranus and Neptune was defined "a timely milestone, fully appropriate for an L class mission". Among the proposed scientific themes, we presented the scientific case of exploring both planets and their satellites in the framework of a single L-class mission and proposed a mission scenario that could allow to achieve this result. In this work we present an updated and more complete discussion of the scientific rationale and of the mission concept for a comparative exploration of the ice giant planets Uranus and Neptune and of their satellite systems with twin spacecraft. The first goal of comparatively studying these two similar yet extremely different systems is to shed new light on the ancient past of the Solar System and on the processes that shaped its formation and evolution. This, in turn, would reveal whether the Solar System and the very diverse extrasolar systems discovered so far all share a common origin or if different environments and mechanisms were responsible for their formation. A space mission to the ice giants would also open up the possibility to use Uranus and Neptune as templates in the study of one of the most abundant type of extrasolar planets in the galaxy. Finally, such a mission would allow a detailed study of the interplanetary and gravitational environments at a range of distances from the Sun poorly covered by direct exploration, improving the constraints on the fundamental theories of gravitation and on the behaviour of the solar wind and the interplanetary magnetic field.

---

[*] Corresponding author; Tel.: +39 0649934414; Fax: +39 0649934383; E-mail: diego.turrini@iaps.inaf.it.

The Comparative Exploration of the Ice Giant Planets with Twin Spacecraft

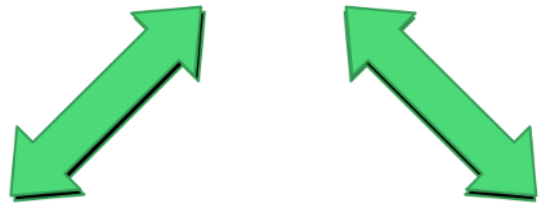

*Figure 1: Scientific themes of ESA Cosmic Vision 2015-2025 program addressed by a space mission to the ice giants Uranus and Neptune.*

## 1. Introduction

The planets of our Solar System are divided into two main classes: the terrestrial planets, populating the inner Solar System, and the giant planets, which dominate the outer Solar System. The giant planets, in turn, can be divided into the gas giants Jupiter and Saturn, whose mass is mostly constituted of H and He, and the ice giants Uranus and Neptune, whose bulk composition is instead dominated by the combination of the astrophysical ices $H_2O$, $NH_3$ and $CH_4$ with metals and silicates. While H and He constitute more than 90% of the masses of the gas giants, they constitute no more than 15-20% of those of the ice giants (Lunine 1993). The terrestrial planets and the gas giants have been extensively studied with ground-based observations and with a large numbers of dedicated space missions. The bulk of the data on the ice giants, on the contrary, has been supplied by the NASA mission Voyager 2, which performed a fly-by of Uranus in 1986 followed by one of Neptune in 1989.

The giant planets likely appeared extremely early in the history of the Solar System, forming across the short time-span when the Sun was still surrounded by a circumstellar disk of gas and dust and therefore predating the terrestrial planets. The role of the giant planets in shaping the formation and evolution of the young Solar System was already recognized in the pioneering works by Oort and Safronov in 1950-1960. In particular, Safronov (1969) suggested that Jupiter's formation would have injected new material, in the form of planetesimals scattered by the gas giant, in the formation regions of Uranus and Neptune. More recently, a revised interpretation of planetary formation, obtained by studying extrasolar planetary systems, gave rise to the idea that the Solar System could have undergone a much more violent evolution than previously imagined (e.g. the Nice Model for the Late Heavy Bombardment, Tsiganis et al. 2005), in which the giant planets played a major role in shaping the current structure of the Solar System.

The scientific case of a space mission to both the ice giants Uranus and Neptune and to their satellite systems and the associated mission concept were first illustrated in the white paper "The ODINUS Mission Concept"[1] submitted to European Space Agency (ESA) in response to its call for white papers[2] for the definition of the scientific themes of the L2 and L3 missions. The purpose of this paper is to provide an updated and expanded discussion, building on the feedbacks the

---

[1] http://odinus.iaps.inaf.it or, on ESA website, http://sci.esa.int/jump.cfm?oid=52030. The ODINUS acronym is derived from the main fields of investigation of the mission concept: Origins, Dynamics and Interiors of the Neptunian and Uranian Systems.
[2] See ESA's website at http://sci.esa.int/Call-WP-L2L3.

The Comparative Exploration of the Ice Giant Planets with Twin Spacecraft

ODINUS white paper received from ESA and the scientific community at large, of this scientific case and of its relevance for advancing our understanding of the ancient past of the Solar System and, more generally, of how planetary systems form and evolve. While we will mainly focus on the open questions that the comparative exploration of the ice giants can address, to better illustrate the challenges presented by performing such a task within a single space mission and the feasibility of the proposed approach, we will also provide a concise but updated description of the ODINUS mission concept, based on the ideas discussed in the white paper and on the results of the following interactions with ESA and the scientific community.

From the perspective of ESA Cosmic Vision 2015-2025 program, the focus of such a mission and of this paper is on the first scientific theme "What are the conditions for planetary formation and the emergence of life?" (see Fig. 1). The study of the formation of the Solar System, however, cannot be separated from that of its present state and of the physical processes that govern its evolution. In discussing the scientific case for a mission to the ice giants, therefore, we will address also the second and third scientific themes of the Cosmic Vision 2015-2025 program, i.e. "How does the Solar System work?" and "What are the fundamental physical laws of the Universe?" (see Fig. 1). In the following we will use these scientific themes of the ESA Cosmic Vision 2015-2025 program as the guideline to discuss the scientific case for a mission to the ice giants and their satellite systems (Sects. 2, 3 and 4). The ODINUS mission concept and the scientific rationale of its twin spacecraft approach will be discussed in Sect. 5 together with the preliminary assessment of its feasibility performed by ESA. Finally, in Sect. 6 we will summarize the outcomes of the selection of the scientific themes for the L2 and L3 missions by ESA, with a focus on the evaluation of the scientific relevance and timeliness of the exploration of the ice giants and on future prospects.

## 2. Theme 1: What are the conditions for planetary formation and the emergence of life?

In this section we will briefly summarize how our understanding of the processes of planetary formation has evolved across the years, discuss their chronological sequence for what concerns the Solar System and highlight how the exploration of Uranus, Neptune and their satellite systems can provide deeper insight and better understanding of the history of the Solar System and how it compares to what we learned from the extrasolar systems discovered to date. It must be noted that the present knowledge on this subject is limited by current observational capabilities and likely supplies only an incomplete view. However, it is not easy to provide a sense of how our knowledge of exoplanets will evolve by the time an L class mission to the ice giants will be launched (currently, no earlier than L4 or 2040). Future space telescopes like ESA M3 Plato and NASA Transiting Exoplanet Survey Satellite (TESS) will explore regions of the phase-space currently unreachable, making it difficult to predict whether the new exoplanets will conform to the partial picture we can draw so far or if we are going to be surprised once more. Concerning the characterization of exoplanets, the James Webb Space Telescope (JWST) will surely provide information on the atmospheric composition of several extrasolar planets but dedicated missions, like ESA M3 mission candidate Exoplanet Characterization Observatory (EChO), for the systematic investigation of atmospheric composition are not currently planned. For further discussion on the subject we refer the readers to Turrini, Nelson & Barbieri (2014) and references therein.

### 2.1 The Evolving View of Planetary Formation: Solar System and Exoplanets

The original view of the set of events and mechanisms that characterize the process of planetary formation (Safronov 1969) was derived from the observation of the Solar System as it is today. This

The Comparative Exploration of the Ice Giant Planets with Twin Spacecraft

brought about the idea that planetary formation was a local, orderly process that produced regular, well-spaced and, above all, stable planetary systems and orbital configurations. However, with the discovery of more and more planetary systems and of free floating planets (Sumi et al. 2011) through ground-based and space-based observations, it is becoming apparent that planetary formation can result in a wide range of outcomes, most of them not necessarily consistent with the picture derived from the observations of the Solar System. The orbital structure of the majority of the discovered planetary systems seems to be strongly affected by planetary migration due to the exchange of angular momentum with the circumstellar disks (see e.g. Papaloizou et al. 2007 and references therein) in which the forming planets are embedded, and by the so-called "Jumping Jupiters" mechanism (Weidenschilling & Marzari 1996; Marzari & Weidenschilling 2002), which invokes multiple planetary encounters, generally after the dispersal of the circumstellar disk, with chaotic exchange of angular momentum between the different planetary bodies involved and the possible ejection of one or more of them.

The growing body of evidence that dynamical and collisional processes, often chaotic and violent, can dramatically influence the evolution of young planetary systems gave rise to the idea that also our Solar System could have undergone the same kind of evolution and represent a "lucky" case in which the end result was a stable and regular planetary system. The most successful attempt to date to apply this kind of evolution to the Solar System was the so-called Nice Model (Gomes et al. 2005; Tsiganis et al. 2005; Morbidelli et al. 2005; Morbidelli et al. 2007; Levison et al. 2011), a Jumping Jupiter scenario formulated to link the event known as the Late Heavy Bombardment (LHB in the following, see e.g. Hartmann et al. 2000 for a review) to a phase of dynamical instability involving all the giant planets. In the Nice Model, the giant planets of the Solar System are postulated to be initially located on a more compact orbital configuration than their present one and to interact with a massive primordial trans-Neptunian region. The gravitational perturbations among the giant planets are initially mitigated by the trans-Neptunian disk, whose population in turn is eroded. Once the trans-Neptunian disk becomes unable to mitigate the effects of the interactions among the giant planets, the orbits of the latter become excited and a series of close encounters takes place. The net result of the ensuing Jumping Jupiters process, in those scenarios that reproduce more closely the present orbital structure of the Solar System, is a small inward migration of Jupiter and marked outward migrations of Saturn, Uranus and Neptune (Tsiganis et al. 2005; Levison et al. 2011).

The importance of the Nice Model lies in the fact that it strongly supports the idea that the giant planets did not form where we see them today or, in other words, that what we observe today is not necessarily a reflection of the Solar System as it was immediately after the end of its formation process. Particularly interesting in the context of the study of Uranus and Neptune is that, in about half the cases considered in the Nice Model scenario, the ice giants swapped their orbits (Tsiganis et al. 2005). The success of the Nice Model in explaining several features of the Solar System opened the road to more extreme scenarios, also based on the migration of the giant planets and the Jumping Jupiters mechanism, either postulating the existence of a now lost fifth giant planet (Nesvorny et al. 2011; Batygin et al. 2012; Nesvorny and Morbidelli 2012) or postulating an earlier phase of migration and chaotic evolution more violent and extreme than the one described in the Nice Model (Walsh et al. 2011). One of the most fascinating aspects of these scenarios is that they all invoke a certain degree of mixing of the solid materials that compose the Solar System. The mixing is generally the larger the more the causing event is located toward the beginning of the Solar System's lifetime. As an example, the "Grand Tack" scenario (Walsh et al. 2011) implies a much stronger remixing than the one that the LHB would cause in the framework of the Nice Model (see e.g. Levison 2009).

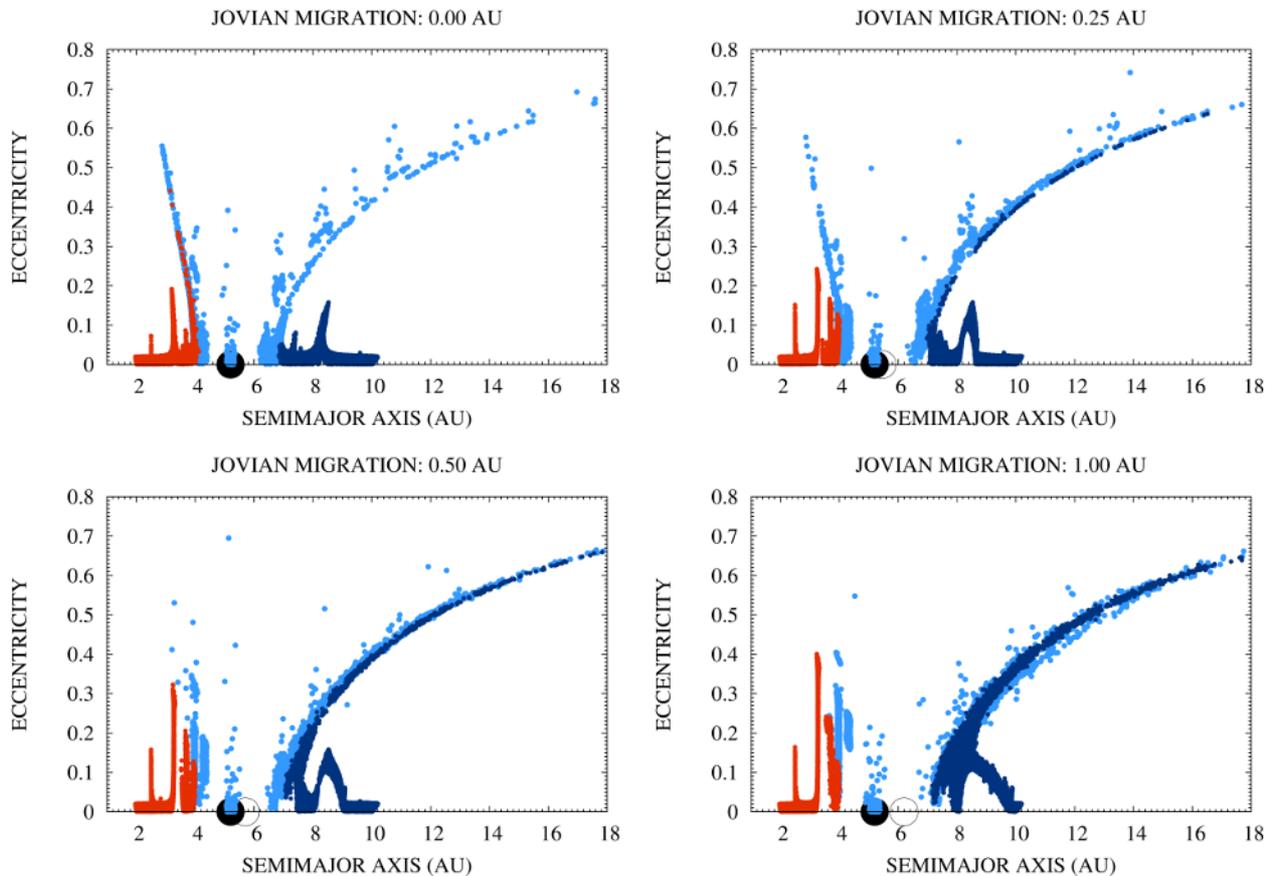

*Figure 2: orbital distribution of the Solar Nebula $2\times10^5$ years after the beginning of the accretion of the nebular gas by Jupiter in the simulations performed by Turrini et al. (2011). The cases considered encompass the classical scenario with no migration (top left), moderate migration (0.25-0.5 au, top right and bottom left) and extensive migration (1 au, bottom right). Planetesimals that formed between 2 and 4 au are indicated in red, those that formed between 4 and 7 au in light blue and those that between 7 and 10 au in dark blue. The open circles are the positions of Jupiter at the beginning of the simulations, the filled ones are the position of Jupiter once fully formed. The excited planetesimals outside 6 au represent the outward flux predicted by Safronov (1969).*

A more or less extensive migration of the giant planets is not required, however, to have a remixing of the solid material in the Solar Nebula. As the pioneering work of Safronov (1969) pointed out, the formation of Jupiter would scatter the planetesimals in its vicinity both inward and outward from its orbit (the "Jovian Early Bombardment" scenario, see Fig. 2 and Turrini et al. 2011, 2012; Coradini et al. 2011; Turrini 2013; Turrini & Svetsov 2014) . In particular, the outward flux of ejected material was postulated by Safronov (1969) to raise the density of solid material in the formation regions of Uranus and Neptune and increase their accretion rate to make it consistent with the lifetime of the Solar Nebula. The inward flux crosses the regions of the terrestrial planets and the asteroid belt, with potentially important implications for the collisional and compositional evolution of the inner Solar System (see Fig. 2 and Weidenschilling 1975, Weidenschilling et al. 2001; Turrini et al. 2011, 2012; Turrini 2013; Turrini & Svetsov 2014). The influence of Jupiter's formation, however, is not limited to the scattering of neighbouring planetesimals: the orbital resonances with the planet would extract planetesimals from more distant regions and put them on orbits crossing those of the other forming giant planets (see Fig. 2 and Weidenschilling et al. 2001; Turrini et al. 2011, 2012; Turrini 2013; Turrini & Svetsov 2014; Turrini, Nelson & Barbieri 2014).

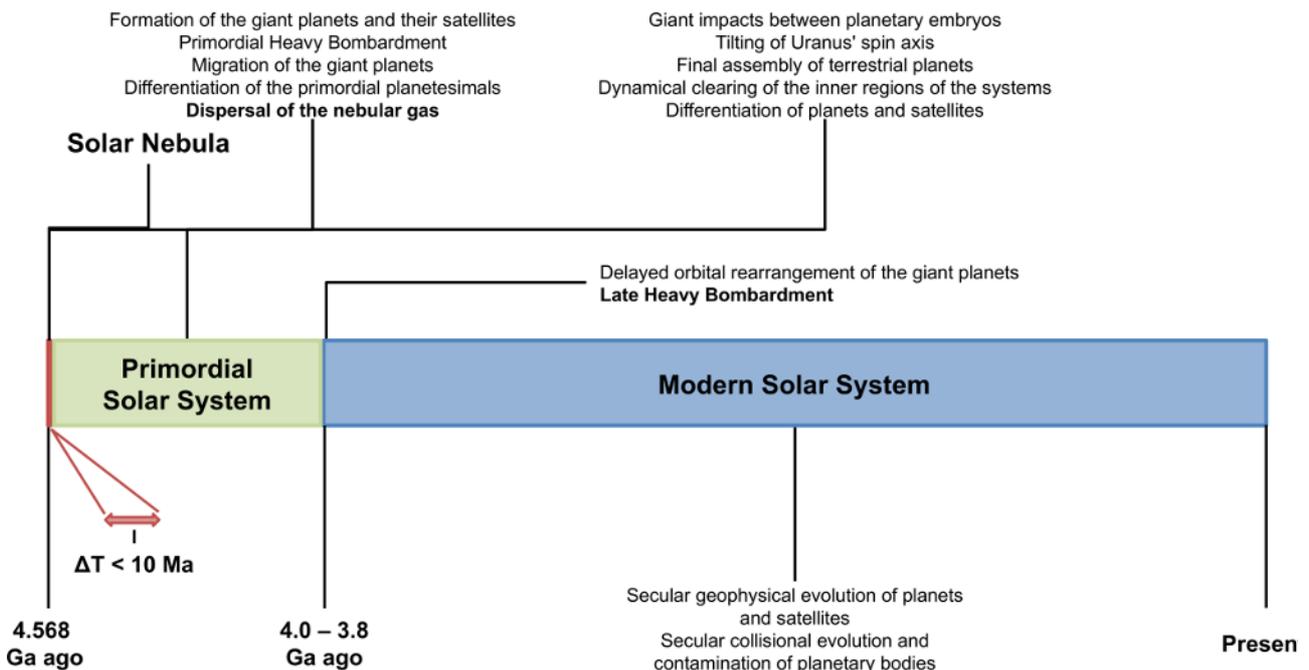

*Figure 3: timeline of the history of the Solar System following the division in three phases (Solar Nebula, Primordial and Modern Solar System) proposed by Coradini et al. (2011). The events marking the transition between the different phases are in bold characters.*

One of the regions affected by the orbital resonances is the asteroid belt (see Fig. 2 and Turrini et al. 2011, 2012; Turrini, Nelson & Barbieri 2014): rocky material can therefore be extracted from the inner Solar System and, as in the original idea from Safronov (1969), possibly be accreted by the forming giant planets (see e.g. Nelson, Turrini and Barbieri 2013 and Turrini, Nelson and Barbieri 2014 for the case of Jupiter) or captured in their circumplanetary disk and incorporated in their satellites.

## 2.2 The Role of Ice Giants in Unveiling the Past of Solar System

As discussed in the previous section, during its history the Solar System went through a series of violent processes that shaped its present structure. The main actors of these processes were the giant planets Jupiter and Saturn. Due to their smaller masses and their likely later formation, Uranus and Neptune were strongly affected by these very same processes together with the rest of the protoplanetary disk. In this section, we will reorganize the events discussed in the Sect. 2.1 in a chronological order and discuss their implications for Uranus and Neptune and their satellite systems. If we follow the description of the history of the Solar System by Coradini et al. (2011), we can divide it into three main phases: the *Solar Nebula*, the *Primordial Solar System* and the *Modern Solar System*. This schematic view of the evolution of the Solar System is summarized in Fig. 3, where we report the main events that took place across the different phases.

### 2.2.1 The Solar Nebula

From the point of view of the giant planets, the Solar Nebula (see Fig. 3) is the period across which they were forming in the circumsolar disk and migrating due to disk-planet interactions. While the gas giants Jupiter and Saturn are forming, the sudden increase of their gravitational perturbations (due to their rapid gas accretion phases) causes a sequence of bombardment events throughout the protoplanetary disk, which Coradini et al. (2011) called the Primordial Heavy

The Comparative Exploration of the Ice Giant Planets with Twin Spacecraft

Bombardment. The prototype of the Primordial Heavy Bombardment is the Jovian Early Bombardment (see Fig. 2, Sect. 2.1 and Turrini et al. 2011, 2012; Turrini 2013; Turrini & Svetsov 2014), triggered by the formation of Jupiter, which was likely the first gas giant to form. The later formation of Saturn would cause a second, similar event, plausibly of lower intensity due to its smaller mass with respect to Jupiter (Coradini et al. 2011). One of the consequences of the Primordial Heavy Bombardment is that, after the formation of the first giant planet, each successive giant planet forms from a more and more evolved and remixed disk, in which the abundances of the various elements and materials are different from the original ones, with implications for the rock/ice ratio and the ratio between different ices in the cores of the giant planets and in the material available for the forming satellites. Measuring the composition of the ice giants and their satellites, and in particular the abundances of noble gases and the isotopic ratios of the different elements, therefore provide a window on the dynamical evolution of the Solar System.

In the classical view of the formation of the Solar System (Safronov 1969), the migration of the giant planets due to their exchange of angular momentum with the circumsolar disk was absent and the main role in reshuffling the protoplanetary disk was played by the Primordial Heavy Bombardment. Recent results in the study of the implications of the Jovian Early Bombardment for the asteroid (4) Vesta (Turrini 2013; Turrini & Svetsov 2014) suggest, however, that this view should be amended and a moderate migration of Jupiter (of the order of 0.25-0.5 au) is required to fit the observational data. As shown in Fig. 2, even such a limited displacement of the giant planet would have implications for the reshuffling of the different materials in the Solar System. In the alternative scenarios we discussed in Sect. 2.1, the proposed extreme migration of the giant planets would have played a more significant role in the reshuffling of the different materials in the Solar System. Specifically, in the "Grand Tack" scenario (Walsh et al. 2011) the giant planets are hypothesized to migrate extensively across the Solar System. Their formation regions, in this case, would be markedly different from those assumed by the classical scenario (both in the case of a moderate migration and in that of a more marked migration like the one shown in Fig. 2) and the composition of their planetary cores would be affected by it.

Part of the planetesimals that the giant planets scatter or excite while forming and migrating would collide with the giant planets themselves, resulting in a "late veneer" of high-Z elements delivered into their atmospheres (Turrini, Nelson & Barbieri 2014). The contribution of high-Z elements provided by this phase of late accretion could have contributed to the super-solar abundances of C, N, S, Ar, Kr and Xe in the atmosphere of Jupiter measured by the probe released by the NASA mission Galileo and those measured in the atmospheres of the other giant planets (see Wong et al. 2008 for a more in-depth discussion of the measured abundances and the proposed causes and Turrini, Nelson & Barbieri 2014). All these remixing events, moreover, affect the source materials, captured in the form of planetesimals by the circumplanetary disks, from which the regular satellites of the giant planets can form (see Coradini et al. 2010 for a review). Depending on the formation time of the relevant giant planet and on the amount of radiogenic sources incorporated in the rocky fraction of the source material, the regular satellites could already differentiate across this phase of the life of the Solar System. Finally, across the Solar Nebula phase a first generation of irregular satellites of the giant planets could have been captured from the protoplanetary disk due to collisions, the effects of gas drag or a combination of the two (see e.g. Mosqueira et al. 2010 for a discussion). This first generation of irregular satellites, however, would not survive the LHB, if the latter is associated with a dynamical instability of the outer Solar System like the one hypothesized by the Nice Model, and a second generation would be created by the LHB itself (Nesvorny et al. 2007).

## *2.2.2. The Primordial Solar System*

Somewhere between the Solar Nebula and the Primordial Solar System phases (see Fig. 3), two events contributed to shape the Uranian and Neptunian satellite systems. One was the giant impact of a planetary embryo with Uranus, suggested to be responsible for its 98° obliquity. As discussed by Coradini et al. (2010) and references therein, it is possible that the original satellite system of the ice giant was destroyed during this event and new satellites formed from the debris of the original ones. The second event was the capture of Triton by Neptune and the following shrinking and circularization of its orbit, which caused the removal of most of the original regular satellites of the ice giant (see e.g. Mosqueira et al. 2010 for a discussion). Across these events and during the first 100 Ma of the life of the Solar System, the giant planets would continue perturbing the planetesimals and planetary embryos residing in the inner and outer Solar System: part of these perturbed bodies (a few per cent in the classical scenario, Guillot & Gladman 2000) would impact againts the giant planets themselves and result in a secular phase of late accretion. The captured mass during this secular accretion appears to be of the same order of magnitude as that delivered by the Primordial Heavy Bombardment (Turrini, Nelson & Barbieri 2014).

Throughout the Primordial Solar System, the Nice Model predicts that the giant planets would still be on different, closer orbits with respect to their present ones. Once the dynamical instability responsible for the LHB takes place, icy planetesimals from what will become the trans-neptunian region are more efficiently excited into high-eccentricity, giant planet-crossing orbits analogous to those of the present-day Centaurs. A fraction of these planetesimals will impact the giant planets (Matter et al. 2009), but their contribution to the late enrichment of their atmospheres is not enough to explain the currently observed abundances (Matter et al. 2009) and is more limited with respect to the one of planetesimals captured at earlier times (Guillot & Gladman 2000; Turrini, Nelson & Barbieri 2014). A fraction of these planetesimals will also impact on the satellites of the giant planets, contributing to their contamination by exogenous material and possibly supplying energy for their late differentiation (Barr & Canup 2010). In particular, Barr & Canup (2010) argue that the LHB could cause the differentiation of Ganymede but not that of Callisto, in agreement with the available data on their internal structures. Another implication of the Nice Model is that any pre-existing population of irregular satellites would be destroyed as a consequence of the close encounters between the giant planets (Tsiganis et al. 2005). However, Nesvorny et al. (2007) showed that three-body effects between the giant planets and the planetesimals during the planetary encounters invoked by the Nice Model would naturally supply a way to re-populate the satellite systems of the giant planets with irregular satellites. It must be noted that these studies are based on the earlier formulation of the Nice Model (Tsiganis et al. 2005; Gomes et al. 2005; Morbidelli et al. 2005) and that the implications of its more recent formulations (Morbidelli et al. 2007; Levison et al. 2011) are still to be addressed. Nevertheless, they show that the evolution of the Solar System across the Primordial Solar System phase could have a non-negligible role in shaping the present-day Uranus and Neptune and their satellite systems.

## *2.2.3 The Modern Solar System*

The Modern Solar System phase starts after the end of the LHB (see Fig. 3) and, differently from the previous two phases, instead of violent processes it is dominated by more regular, secular ones. Moreover, the population of small bodies in the outer Solar System is significantly smaller than that at earlier times, so that collisional processes are less intense than before. Most of the information that we can gather through crater counting on the surface of the satellites of the giant planets refers to this long, more quiescent phase, especially if the satellites are still geophysically active and undergo resurfacing, as it appears to be the case for Triton (see Schubert et al. 2010 for a discussion). In the case of geophysically active satellites, moreover, the surface features and composition supply us information on their more recent internal state, i.e. they again give us insights into the processes that acted across the Modern Solar System phase. Depending on the degree of geophysical activity and the flux of impactors (both planetocentric, i.e. other satellites, and heliocentric, e.g. comets and Centaurs), the surfaces of the satellites can be contaminated to various degrees by exogenous materials (see e.g. Mosqueira et al. 2010; Schubert et al. 2010 for a discussion), an effect that has to be taken into account while interpreting spectral data, which typically probe just the first few mm of the satellite surfaces. Across the Modern Solar System, moreover, the secular effects of space-weathering due to various exogenous sources (e.g. solar wind, magnetospheric plasma, cosmic rays) contribute to the surface evolution of the satellites in ways that are still poorly quantified or even understood.

## *2.3 The exploration of Uranus and Neptune and the history of the Solar System*

As Sects. 2.1 and 2.2 highlight, our view of the processes of planetary formation and of the evolution of the Solar System has greatly changed over the last twenty years but most of the new ideas are in the process of growing to full maturity or need new observational data to test them against. The comparative study of Uranus and Neptune and their satellite systems will enable outstanding problems to be addressed, as the ice giants were affected more than other planets by the violent processes that sculpted the early Solar System and yet they are the least explored and more mysterious of the giant planets. In particular, the exploration of Uranus and Neptune and of their satellite systems allows probing those phases of the life of the Solar System preceding the formation of the terrestrial planets, which completed their assembly only after a few $10^7$ years (see Fig. 3).

The primary information that a mission to Uranus and Neptune should gather to investigate the history of the Solar System are:
- What are the atmospheric composition and enrichment with respect to the solar abundances of the two planets?
- What are the bulk densities and the masses of the ice giants and their satellites?
- What are the interior structures and density profiles of the ice giants and their satellites?
- What is the surface composition of the regular and irregular satellites?
- Which satellites are fully or partially differentiated and which ones are undifferentiated?

Using these data, the open questions that such a mission can help to answer are:
- When and where did the planets form? Did they migrate? If so, how much? Did Uranus and Neptune swap their positions as hypothesized by the Nice Model?
- What is the ice-to-rock ratio of the cores of ice giants and of their satellites? How much "non-local" material was available to them when they formed? Where did this "non-local" material originated from?
- Are the satellites of Uranus primordial or did they reform after the planet tilted its spin axis? What were the effects of the capture of Triton for the Neptunian satellites?
- Where did the irregular satellites originate? Can they be used to constrain the dynamical evolution of the ice giants?

Note that the questions and the related measurements here reported do not aim to address all the possible investigations that a mission to the ice giants could perform, but focus on the primary driver of the proposed mission concept, i.e. the study of the past history of the Solar System. A discussion of several other measurements and studies that such a mission will allow, albeit non-exaustive, is provided is Sect. 3.

The Comparative Exploration of the Ice Giant Planets with Twin Spacecraft

## *2.4 Uranus and Neptune as templates for the extrasolar planets*

As we detailed in Sect. 2.3, a mission to the ice giants has the potential to provide precious information on the history of our Solar System and on the processes that shaped its formation and evolution. Moreover, the study of the ice giants is important also to gain deeper insight on one of the most abundant classes of extrasolar planets according to the observational sample to date. Based on the data supplied by the NASA mission Kepler once corrected for selection effects, about one star out of five in our galaxy should possess at least a Neptune-like planet (Fressin et al. 2013). While there is a growing amount of efforts devoted to the characterization of the atmospheric composition of giant exoplanets with ground-based or space-based facilities (see e.g. Turrini, Nelson & Barbieri 2014 and references therein), the only observational ground-truth we possess on this class of planets, especially from the point of view of their interior, is represented by the observations performed by Voyager 2 during its fly-bys of Uranus in 1986 and of Neptune in 1989 and from ground-based observations that, however, cannot supply the same coverage over all phase angles and observing geometries and cannot achieve the same spatial resolution as the one obtained from a spacecraft.

It is important to point out that the Neptune-like candidates discovered by Kepler so far have orbital periods of less than about 1 year: they are therefore characterized by orbits between one and two orders of magnitude closer to their host stars than they solar counterparts Uranus and Neptune. Because of this, these exoplanets are generally classified as "warm" or "hot" Neptunes depending on their atmospheric temperatures. Their atmospheric composition and meteorology are expected to be extremely different from those of the ice giants in the Solar System; currently, however, data are available only for the atmospheric composition of one exo-Neptune (source: www.exoplanet.eu), GJ 436 b (Madhusudhan & Seager 2011). Nevertheless, Uranus and Neptune are the only examples of this class of planets within the reach of a space mission and can represent the templates to interpret the data that present and future missions, devoted to the discovery and characterization of exoplanets, will gather. From this point of view, it is particularly important to characterize both ice giants in the Solar System and not just one of them, as we presently don't know whether extreme obliquities like that of Uranus are common or not from a galactic perspective. The study of Uranus and Neptune can therefore provide a key to identify similar configurations in other planetary systems and properly interpret them.

## 3. Theme 2: How does the Solar System work?

A mission devoted to exploring the ice giants and their satellites to unveil the history of the Solar System would gather a wealth of data on the present status of the Uranian and Neptunian systems. The collected data would enable a more complete understanding of how the surfaces and interiors of icy satellites evolve so far from the Sun. Moreover, the coupled investigation of Uranus and Neptune, so similar and yet so different, would provide fundamental new insights into the cause of their different atmospheric and thermal behaviours.

## *3.1 Atmospheres of Uranus and Neptune*

The Herschel observations of Uranus and Neptune (Feuchtgruber et al., 2013) confirmed that the ice giants have a remarkably similar D/H content ($4.4\pm0.4\times10^{-5}$ and $4.1\pm0.4\times10^{-5}$ respectively), suggesting a common source of icy planetesimals in the protoplanetary disk. Further insight on the conditions of the disk in its outer regions can be derived from the relative enrichment (with respect to the Solar values) of C, N, S and O, by determination of the abundances of the corresponding reduced forms. At the current date, methane is still the only of these reduced forms that has been

The Comparative Exploration of the Ice Giant Planets with Twin Spacecraft

directly detected in both ice giants (e.g.: Baines et al., 1994)[3]. Recent analyses by Karkoschka and Tomasko (2009) and Tice et al., (2013) indicate that the methane mixing ratio varies with latitude. An extensive investigation of the minor gases in the atmospheres of the icy giants (with a special attention to their horizontal and vertical variations) is therefore extremely urgent to ultimately characterize the emergence of our Solar System.

The post-Voyager 2 observations of Uranus by ground-based and space telescopes revealed a progressive increase of meteorological activity (cloud and dark spots occurrence) in the proximity of Northern Spring equinox (see, e.g. Sromovsky et al., 2012). While this evolution is undoubtedly related to the extreme obliquity of the planet, the relative roles of solar illumination and internal heating (and its possible variations) remain to be assessed by detailed studies at high spatial resolution. Even more important, spacecraft infrared observations will provide an extensive coverage of the night hemisphere. The possibility of comparing the atmospheric behaviour of Uranus with the extremely dynamic meteorology of Neptune provides a unique opportunity to gain insights on the response of thick atmospheres to time-variable forcing, representing therefore a new area of tests for future atmospheric global circulation models, in conditions not found in terrestrial planets or gas giants.

Uranus zonal winds are currently characterized by moderately retrograde values (-50 m s$^{-1}$) at the equator that progressively become prograde, to reach a maximum value of 200 m s$^{-1}$ at 50N (Sromovsky et al., 2012). On Neptune, a similar pattern is observed, but the absolute speed values are strongly amplified, to reach – despite the limited solar energy input – some of the most extreme values (400 m s$^{-1}$ or more) observed in the Solar System (Martin et al., 2012). Wind speed fields are the most immediate proxy for atmospheric circulation and their modeling can provide constraints on very general properties of the atmosphere, such as the extent of deep convection (Suomi et al., 1991). While ground based observers have considerably expanded the results of Voyager 2, an extensive, long-term, and high spatial resolution cloud tracking campaign remains essential to study the ultimate causes of these extreme phenomena.

Patterns of zonal winds of ice giants as revealed by available data are also noteworthy for their lack of coherence (variation of absolute values, high dispersions and differences in results from different spectral bands) once compared to the Jupiter and Saturn cases (see Hammel et al., 2001, Hammel et al., 2005 for Uranus, Sromovsky et al., 1993 and Fitzpatrick et al., 2014 for Neptune). The assessment of the relative role of different phenomena (such as vertical wind shear, transient clouds due to dynamically driven sublimation and condensation, temporal variations on different time scales) will highly benefit from the long-term, high spatial resolution monitoring.

Neptune shows an unexpected temperature of 750 K in its stratosphere (Broadfoot et al., 1989) that cannot be justified by the small solar UV flux available at that heliocentric distance. More complex mechanisms – such as energy exchange with magnetospheric ions (Soderlund et al. 2013)– shall become predominant in these regions. Uranus, on the other hand, offers unique magnetospheric geometries because of its high obliquity and strong inclination of magnetic axis (see also Sect. 3.3).

## 3.2 *The satellites of Uranus and Neptune*

The geological history and the composition of the satellites of Uranus and Neptune are poorly known due to the limited resolution and surface coverage of the Voyager 2 observations. The Uranian satellites Ariel and Miranda showed a complex surface geology, dominated by extensional tectonic structures plausibly linked to their thermal and internal evolution (see Prockter et al. 2010 and references therein). Umbriel appeared featureless and dark, but the analysis of the images suggests an ancient tectonic system (see Prockter et al. 2010 and references therein). Little is known about Titania and Oberon, as the resolution of the images taken by Voyager 2 was not enough to distinguish tectonic features, but their surfaces both appeared to be affected by the presence of dark material. The partial coverage of the surface of Triton revealed one of the youngest surfaces of the Solar System, suggesting the satellite is possibly more active than Europa (see Schubert et al. 2010 and references therein). Notwithstanding this, the surface of Triton showed a variety of cryovolcanic, tectonic and atmospheric features and processes (see Prockter et al. 2010 and references therein). The improved mapping of these satellites, both in terms of coverage and resolution, would enable much improved measurements of their crater records and their surface morphologies, which in turn would provide a deeper insight into their past collisional and geophysical histories.

---

[3] Detections of water vapour (e.g.: Feuchtgruber, H. et al. 1994, Lellouch et al., 2010) are interpreted as due to exogenic stratospheric gases and are therefore not relevant to constraint the bulk composition of the two planets.

From the point of view of their surface composition, the Uranian satellites are characterized by the presence of crystalline $H_2O$ ice (see Dalton et al. 2010 and references therein). The spectral features of Ariel, Umbriel and Titania showed also the presence of $CO_2$ ice, which however should be unstable over timescales of the order of the life of the Solar System, while $CO_2$ ice was not observed on Oberon (see Grundy et al. 2006; Dalton et al. 2010 and references therein). In the case of Miranda, the possible presence of ammonia hydrate was observed but both the presence of the spectral band and its interpretation are to be confirmed (see Dalton et al. 2010 and references therein). The confirmation of the presence of ammonia would be of great importance due to its anti-freezing role in the satellite interiors. The spectra of Triton possess the absorption bands of five ices: $N_2$, $CH_4$, CO, $CO_2$, and $H_2O$ (Dalton et al. 2010). The detection of the HCN ice band has been reported, which could imply the presence of more complex materials of astrobiological interest (see Dalton et al. 2010 and references therein). Triton also possesses a tenuous atmosphere mainly composed of $N_2$ and CO, which undergoes seasonal cycles of sublimation and re-condensation (see Dalton et al. 2010 and references therein). Images taken by Voyager 2 revealed active geyser-like vents on the surface of Triton, indicating that the satellite is still geologically active (even if at present it is not tidally heated, see Schubert et al. 2010 and references therein) and, similarly to the Saturnian satellite Enceladus (Spencer et al. 2009 and references therein), possesses liquid water in its interior, sharing its astrobiological potential as a possible sub-surface habitable habitat.

Both Uranus and Neptune possess a family of irregular satellites. Neptune, in particular, possesses the largest irregular satellite (not counting Triton) in the Solar System, i.e. Nereid. Aside their estimated sizes and the fact that observational data suggest they might be more abundant than those of Jupiter and Saturn (Jewitt and Haghighipour 2007), almost nothing is known of these bodies. The collisional evolution of the irregular satellites results in the secular production of dust, as supported by observational data in the Jovian and Saturnian systems (Krivov et al. 2002, Verbiscer et al. 2009). Depending on their sizes, the non-gravitational forces can either strip away the dust particles from their planetocentric orbits or cause them to spiral inward and impact with the regular satellites or the planets (see Schubert et al. 2010 and references therein; see also Tosi et al. 2010 and Tamayo et al. 2011 for the specific case of the Saturnian system). In the case of the Uranian system, Tamayo et al. (2013) recently showed that the latter effect would affect the surface of the four outermost regular satellites (due to the dynamical instability caused by the obliquity of Uranus) and could explain the increasing trend of leading-trailing color asymmetries of the hemispheres of the satellites with planetocentric distance observed by Buratti and Mosher (1991). The study of the irregular satellites would therefore constrain the origin of the dark material, and likely other contaminants, observed on some of the Uranian satellites and discriminate whether it originated from the irregular satellites or it was the result of local (e.g. the interaction with the magnetosphere) or endogenous processes.

### *3.3 Magnetosphere-Exosphere-Ionosphere coupling in the Uranian and Neptunian systems*

Neptune and Uranus have strong non-axial multipolar magnetic field components compared with the axial dipole component (Connerney et al., 1991; Herbert, 2009). The magnetic fields of both planets are generated in the deep, electrically conducting regions of their interiors, i.e. in electrolyte layers composed of water, methane and ammonia (Hubbard et al., 1991; Nellis et al., 1997) or superionic water (Redmer et al., 2011). Numerous modelling efforts have shown that the mechanism of a dynamo operating in a thin shell surrounding a stably-stratified fluid interior produces magnetic field morphologies similar to those of Uranus and Neptune (Hubbard et al., 1995; Holme and Bloxham, 1996; Stanley and Bloxham, 2006). In addition, Gómez-Pérez and Heimpel (2007) showed that weakly dipolar and strongly tilted dynamo fields are stable in the presence of strong zonal circulation and when the flow has a dominant effect over the magnetic fields. Guervilly et al. (2012) proposed that if some mechanism is able to transport angular momentum from the surface down to the deep, fully conducting region then the zonal motions may influence the generation of the magnetic field. Such zonal jets at the giant planets may exert, by viscous or electromagnetic coupling, an external forcing at the top of the deeper conducting envelope. The model by Guervilly et al. (2012) assumes an idealized one-way coupling between the outer and deep regions, assuming a constant (throughout the whole modeled layer) conductivity and ignoring the back reaction of the deep layer onto the outer layer. In order to assess the role of zonal winds in the generation and topology of the magnetic fields of Uranus and Neptune, determination of the compressibility of the layers, of the radial profiles of the electrical conductivity, of the viscosity and of the viscous coupling between electrically insulating and conducting regions, is necessary. These quantities cannot be estimated from direct measurements. Nonetheless, the proposed mission concept can provide key constraints for further modeling efforts devoted to characterizing the longitudinal profile of zonal winds at the cloud top and their possible secular variations (by means of visible and IR imaging), the magnetic field and the gravitational field. Namely, the orbits of the two spacecraft can be optimized to allow determination of gravity fields at least up to order 12, to assess the scale height of exponential decay of zonal winds along the rotation axis (Kaspi et al, 2010), which constrain the degree of dynamic coupling between surface and interior.

The highly non-symmetric internal magnetic fields of Uranus and Neptune (Ness et al. 1986, 1989; Connerey et al. 1991; Guervilly et al., 2012), coupled with the relatively fast rotation and the unusual inclination of the rotation axes from the orbital planes, imply that their magnetospheres are subject to drastic geometrical variations on both diurnal and seasonal timescales. The relative orientations of the planetary spin and their magnetic dipole axes and the direction of the solar wind flow determine the configuration of each magnetosphere and, consequently, the plasma dynamics in these regions.

Due to the planet's large obliquity, Uranus' asymmetric magnetosphere varies from a pole-on to a orthogonal configuration during a Uranian year (84 Earth years) and changes from an "open" to a "closed" configuration during a Uranian day. At solstice (when Uranus' magnetic dipole simply rotates around the vector of the direction of the solar wind flow) plasma motions due to the rotation of the planet and by the solar wind are effectively decoupled (Selesnick and Richardson, 1986; Vasyliunas, 1986). Moreover, the Voyager 2 plasma observations showed that when the Uranus dipole field is oppositely directed to the interplanetary field, injection events to the inner magnetosphere (likely driven by reconnection every planetary rotation period) are present (Sittler et al., 1987). The time-dependent modulation of the magnetic reconnection sites, the details of the solar wind plasma entry in the inner magnetosphere of Uranus and the properties of the plasma precipitation to the planet's exosphere and ionosphere are unknown. Models indicate that Uranus' ionosphere is dominated by $H^+$ at higher altitudes and $H_3^+$ lower down (Capone et al., 1977; Chandler and Waite, 1986; Majeed et al., 2004), produced by either energetic particle precipitation or solar ultraviolet (UV) radiation. Our current knowledge of the aurora of Uranus is limited since it is based only on 1) a spatially resolved observation of the UV aurora (by the Ultraviolet Spectrograph data on board Voyager 2, Herbert 2009), 2) observations of the FUV and IR aurora with the Hubble Space Telescope (Ballester, 1998), and 3) observations from ground-based telescopes (e.g., Trafton et al., 1999). The details of the solar wind plasma interaction with the planet's exosphere, ionosphere and upper atmosphere (possibly through charge exchange, atmospheric sputtering, pick-up by the local field), the seasonal and diurnal variation of the efficiency of each mechanism as well as the total energy balance (deposition/loss) due to magnetosphere-exosphere-ionosphere coupling are unknown. Since the exact mechanism providing the required additional heating of the upper atmosphere of Uranus is also unknown, new in situ plasma and energetic neutral particles observations could become of particular importance to determine whether the extent to which plasma precipitation to the exosphere has a key role in this context. The magnetospheric interaction with the Uranian moons and rings can be studied through in situ measurements of magnetic field, charged particles, and energetic neutrals emitted from the surfaces. Finally, remote imaging of charge exchange energetic neutral atoms (ENAs) would offer a unique opportunity to monitor the plasma circulation where moons and/or Uranus' exosphere are present.

Neptune's magnetic field (Ness et al., 1989; Connerey et al. 1991) has a complex geometry that includes relatively large contributions from localized sources or higher order magnetic multipoles, or both, yet not well determined (Ness et al. 1989). Neptune is a relatively weak source of auroral emissions at UV and radio wavelengths (Broadfoot et al., 1989; Bishop et al., 1995; Zarka et al., 1995). Although this non-observation does not rule out an active magnetosphere per se, it ruled out processes similar to those associated with the aurora observed at Uranus. Whereas the plasma in the magnetosphere of Uranus has a relatively low density and is thought to be primarily of solar-wind origin, at Neptune, the distribution of plasma is generally interpreted as indicating that Triton is a major source (Krimigis et al., 1989; Mauk et al., 1991, 1994; Belcher et al., 1989; Richardson et al., 1991). Escape of neutral hydrogen and nitrogen from Triton maintains a large neutral cloud (Triton torus) that is believed to be source of neutral hydrogen and nitrogen (Decker and Cheng, 1994). The escape of neutrals from Triton could be an additional plasma source for Neptune's magnetosphere (through ionization). Our knowledge of the plasma dynamics in the magnetosphere of Neptune as well as on the neutral particles production in Triton's atmosphere is limited. New in situ plasma and energetic neutral particles observations focused on Triton's region can provide important information on the role of the combined effects of photoionization, electron impact ionization, and charge exchange in the context of the coupling of a complex asymmetric planetary magnetosphere with a satellite exosphere at large distances from the Sun.

### *3.4 Planetary and satellite interiors*

The available constraints on interior models of Uranus and Neptune are limited. The gravitational harmonics of these planets have been measured only up to fourth degree ($J_2$, $J_4$), and the planetary shapes and rotation periods are not well known (see e.g. Helled et al. 2011 and references therein). The response coefficients of Uranus and Neptune suggest that the latter is less centrally condensed than the former (De Pater and Lissauer 2010).

The thermal structures of these planets are also intriguing (see e.g. Helled et al. 2011 and references therein). Uranus stands uniquely among the outer planets for the extremely low value

The Comparative Exploration of the Ice Giant Planets with Twin Spacecraft

(0.042±0.047 W m$^{-2}$) of its internal energy flux (Pearl et al., 1990). This figure sharply contrasts with Neptune, where Voyager 2 determined a value of 0.433±0.046 W m$^{-2}$ (Pearl et al., 1991). The two ice giants must therefore differ in their internal structure, heat transport mechanisms, and/or in their formation history. Substantial differences in internal structures are suggested by the analysis of available gravitational data for the two planets (Podolak et al., 1995). Namely, the Uranus gravity data are compatible with layered convection in the shell, which inhibits the transport of heat. Alternative views call – among the others – for a later formation age of Neptune (Gudkova et al., 1988). Consequently, heat fluxes represent, along with gravity, magnetic data and wind fields (Soderlund et al. 2013), the key experimental constraints to characterize the interior of Uranus and Neptune and their evolution.

The information on the interior structure of the satellites of Uranus and Neptune is even more limited and is mostly derived from their average densities, which are used to infer the rock-to-ice ratios, and their surface geology, which suggests that across their lives they possessed partially or completely molten interiors (De Pater and Lissauer 2010). As a consequence, the data that can be collected by a mission to the ice giants on their interiors will play an important role in filling up this gap in our understanding of the icy satellites in the outer Solar System.

Gravity data can indeed be used to constrain the internal structure and composition of the planets. The gravitational potential due to a body with rotational symmetry can be represented by an harmonic expansion of the type

$$U = \frac{GM}{r}\left(1 - \sum \left(\frac{a}{r}\right)^{2n} J_{2n} P_{2n}(\cos\theta)\right) + \frac{1}{2}\omega^2 r^2 \cos^2\theta \ ,$$

see e.g. Helled et al. 2011, where r, θ, φ are spherical coordinates, G the Newtonian gravitational constant, M the mass of the primary body and ω its rotational angular velocity. The specific potential depends on the zonal coefficients $J_{2n}$. Such deviations of the primary body gravitational field from the spherical symmetry (due to its rotational state and internal structure and composition) perturb the orbit of the spacecraft and can be extracted via a precise orbit determination and parameter estimation procedure from the tracking data (usually range and range-rate in a typical radio science experiment). Fundamental to this objective is a proper modelling of the spacecraft dynamics, both gravitational and non-gravitational. This could be non-trivial in case of a complex spacecraft (the ideal would be a test mass) and – in selected cases – could require also the use of an on-board accelerometer (Iafolla et al., 2010). In the case of Uranus, measurements of the precession of its elliptical rings should add to the list of observables. Of course, this investigation of the internal structure of the primaries can be also extended to their satellites. Indeed, selected fly-bys of the satellites will allow for the determination of their gravitational coefficient and, at least, of their lowest-degree multipoles. The set of estimated parameters could include also the masses of the planets or of their satellites (see the right-hand side of the previous equation). This is a measurement that is difficult to perform remotely, and a direct result of having a probe orbiting the various bodies of the system.

An alternate and complementary method to probe the internal structures of Uranus and Neptune consists of using seismic techniques that were developed for the Sun (helioseismology, see e.g. Goldreich & Keeley 1977), then successfully applied to stars with the ESA space mission CoRoT and the NASA space mission Kepler (Michel et al. 2008, Borucki 2009), and tested on Jupiter (Gaulme et al. 2011). Seismology consists of identifying the acoustic eigen-modes, whose frequency distribution reflects the inner sound speed profile. The main advantage of seismic methods with respect to gravity moments is that waves propagate down to the central region of the planet, while gravitational moments are mainly sensitive to the external 20% of the planetary radius. The second advantage is that the inversion problem is not model dependent, neither on the equation of state or on the abundances that we want to measure. As regards Uranus and Neptune, the difference in internal energy flux should appear as a difference in the amplitude of acoustic modes. Moreover, a by-product of the seismological observations is the map of the wind fields in the atmospheres of the giant planets (Schmider et al. 2009; Murphy et al. 2012), which as mentioned previously provides additional constraints on the interior structure of the planets themselves (Soderlund et a. 2013). As for helioseismology, two approaches may be used to perform such seismic measurements, either with Doppler spectro-imaging (e.g. Schmider et al. 2007), or visible photometry (Gaulme & Mosser 2005). A dedicated study needs to be conducted to determine whether a seismological investigation is feasible in the framework of the mission concept described in Sect. 5 and, in case, which method is the most appropriate for these two planets.

### *3.5 Heliosphere science*

During the mission cruise phase, it will be possible to obtain important information on the interplanetary medium properties at different distances from the Sun as well as on the heliospheric structure and its interactions with the interstellar medium. Although there is plenty of information on how solar wind and coronal mass ejections interact with the interplanetary medium at 1 au from the Sun, little is known on how this interaction works at larger distances. The measurements of the interplanetary magnetic field fluctuations and plasma densities variations at different distances from the Sun, such as those that a mission to the ice giants would allow, can provide information for understanding the origin of turbulence in the solar wind and its evolution from its source to the heliopause. A mission to the ice giants, therefore, would give an opportunity to obtain constraints for the processes of energy transfer in different regions of the solar system and to understand the mechanisms of the energy dissipation.

In order to answer a series of fundamental questions concerning the particle acceleration in the Solar System, the galactic cosmic ray modulation and the plasma/planetary bodies interaction, it is important to have knowledge of the overall structure of the heliosphere. Prevailing models of the shape of the heliosphere suggest a cometary-type interaction with a possible bow shock and/or heliopause, heliosheath, and termination shock (Axford, 1973; Fichtner et al., 2000). However, recent energetic neutral atom images obtained by the Ion and Neutral Camera (INCA) onboard the NASA spacecraft Cassini did not conform to these models (Krimigis et al., 2009). Specifically, the map obtained by Cassini/INCA revealed a broad belt of energetic protons with non-thermal pressure comparable to that of the local interstellar magnetic field (Krimigis et al., 2009). In October 2008, the NASA mission Interstellar Boundary Explorer (IBEX) was launched with energetic neutral atom cameras specifically designed to map the heliospheric boundary at lower (<6 keV) energies (McComas et al., 2009; Funsten et al., 2009). Both IBEX and INCA identified in the energetic neutral atom images dominant topological features (ribbon or belt) that can be explained on the basis of a model that considers an energetic neutral atom-inferred non-thermal proton pressure filling the heliosheath from the termination shock to the heliopause (Krimigis et al., 2009).

The ENA imaging is a promising technique for remote imaging of the heliospheric boundary. Hydrogen ENAs are generated in the heliosheath through charge-exchange between the shocked solar wind protons and the cold neutral interstellar hydrogen gas. The shocked protons in this region are mostly isotropic and some fraction of the resulting ENAs will propagate radially inwards, unimpeded by the interplanetary magnetic field (Hsieh et al., 1992; Gruntman et al., 2001). Synchronized ENA observations with the dual spacecraft of the proposed mission concept will provide a mapping of the shocked solar wind protons, and will reveal information on the heliosheath structure and the properties of the complex interstellar interaction. The proposed measurements of the heliosheath structure, to be performed from the Uranus and Neptune orbits, are required for the achievement of just additional science objectives. However, such measurements could still be of significant interest, since they would complement the IBEX observations extending them to a different Solar cycle and possibly with a better angular resolution, since the spacecraft will be closer to the heliopause. Moreover, in case the spacecraft arrives when Uranus is in the heliotail's side, the expected angular separation between primary and secondary oxygen populations will be higher than the one at 1 au. As a result, these two populations could be discriminated with higher accuracy than in the Earth's orbit case (McComas et al., 2009; Möbius et al., 2009). Different models of interaction of the solar wind with the interstellar medium could be constrained by ENAs observations in a wide energy range comprising IBEX and INCA range. Different vantage points for ENA imaging could be useful to reconstruct the ENA generation geometries.

The design of the ENA cameras is intended to meet the requirements for measuring the ENAs generated in the heliosheath. Combined ENA and magnetic field measurements at the orbits of Uranus and Neptune will provide complementary information (to the one obtained from the Earth's orbit) for addressing the question whether the interaction of the heliosphere with the interstellar magnetic field takes place at the termination shock or at the heliopause.

## 4. Theme 3: What are the fundamental physical laws of the Universe?

Since the early days of interplanetary exploration missions, spacecraft have been used as (nearly) test masses to probe the gravitational machinery of Solar System and, more in general, as a test for fundamental physics. Though general relativity is currently regarded as a very effective description of gravitational phenomena, having passed all the experimental tests (both in the weak- and strong-field regimes, see e.g. Will 2006) so far, it is challenged by theoretical (e.g. Grand Unification, Strings) scenarios (e.g., Damour et al. 2002) and by cosmological findings (Turyshev, 2008). Stringent tests of general relativity have been obtained in the past by studying the motion of spacecraft during their cruise phase, as well as the propagation of electromagnetic waves between spacecraft and Earth (see e.g. Bertotti et al. 2003 for the measurement of the Shapiro time delay and the corresponding improved bound on post-Newtonian parameter $\gamma$). In this respect, the spacecraft are considered as test mass subject (mainly) to the gravitational attraction of Solar System bodies. Well-established equations of motions can then be tested against the experimental data, in order to place strong constraints on possible deviations from what is predicted by general relativity. At the same time, the spacecraft – acting as a virtual bouncing point for microwave pulses – enables a precise measurement of the propagation of electromagnetic waves between Earth and spacecraft (e.g., Shapiro time delay). Being very effective in the past in ruling out possibilities of "exotic physics" (i.e., the so-called "Pioneer Anomaly", see Anderson et al. 1998b), such tests could be used in the future to further pursue experiments in this way. The very-weak-field environment of the more external regions of the Solar System is particularly interesting, in that "exotic" phenomenology such as MOND could be probed (see e.g. Famaey & McGaugh 2012). While it could be possible to replicate the Cassini test for the measurement of $\gamma$, the most interesting possibility offered by the mission will be the opportunity of testing the gravitational interaction at a scale of distances at which few precision measurements are available. Since the standard scenario predicts nothing new at these scales, an eventual signal that could be clearly traceable to a gravitational origin will be a strong candidate of new phenomenology. This possibility however depends on the availability of a very stable reference point given by the spacecraft itself. This implies a strong reduction (or knowledge at the same level of accuracy) of all non-gravitational dynamics. These tests would help extend the scale at which precision information on gravitational dynamics is available; this will contribute to bridging the "local" scale (in which precise measurements on gravitational dynamics are available) to more "global" scales (subject to puzzling phenomenology as dark matter and dark energy).

The Comparative Exploration of the Ice Giant Planets with Twin Spacecraft

The experimental setup needed to perform the previous tests also allows for constraining the amount of non-luminous matter remnant of Solar System formation (e.g., the trans-Neptunian region), as well as the presence of some form of dark matter that could be trapped in a halo around the Sun (Anderson et al. 1989; Anderson et al. 1995). At least two approaches for placing constraints to such an amount of matter can be considered, and in fact have been used with regard to Pioneer 10, Pioneer 11 and Voyager 2 trajectories.

The first assumes a spherically symmetric matter distribution around the Sun, and estimates its gravitational perturbations on bodies outside the distribution (as — case by case — Jupiter, Uranus and Neptune) using range points obtained during flybys. Anderson et al. (1995) obtained the limits of $0.32 \pm 0.49$ for Uranus and $-1.9 \pm 1.8$ for Neptune respectively, both results being in units of $10^{-6}$ $M_\odot$ respectively[4]. The negative sign in Neptune's result is interesting: it may point to a non-spherically symmetric mass distribution inside Neptune's orbit.

The second approach considered ten years of Pioneer 10 trajectory inside what has been supposed to be the trans-Neptunian region; bounds on the density of small-size particles have been obtained from the lack of detectable damage to the spacecraft (namely to the propellant tank). For example, having parameterized the density distribution of Kuiper Belt particles with $n(r) = n_0 r^{-\gamma}$, and taking a large $\gamma$, a bound of $M_\oplus/3$ for $\rho < 0.4$ g cm$^{-3}$ and of $M_\oplus/10$ for $\rho < 0.133$ g cm$^{-3}$, $\rho$ being the particles density, has been obtained in Anderson et al. (1998a).

Such approaches could well be applied to a mission towards the two ice giants, to place further constraints on non-luminous matter. In general, precisely reconstructing the orbit of the probe(s) during the entire cruise will enable a possible repetition of this test at various distances from the Sun. An advantage of a orbiter at the ice giants, with respect to previous measurements, is that the former would provide for a rather long series of measurements instead of the few ones (one for each fly-by pass) reported in the past.

We can notice that this type of experiment, performed instead when the spacecraft is in orbit around one of the two planets, implies an improvement of the corresponding planet ephemerides. Since in the current best implementation of Solar System ephemerides (see e.g. Folkner et al., 2008) the orbits of Uranus and Neptune are not so well determined as the ones of more inner bodies, due in particular to the lack of recent spacecraft tracking[5], any further data in this direction will help to enhance the ephemerides themselves.

## 5. The ODINUS mission concept and the scientific rationale of the twin spacecraft approach

The approach proposed to ESA in the white paper "The ODINUS Mission Concept" was to use a set of twin spacecraft, each to be placed in orbit around one of the two ice giant planets (see Fig. 4). The traditional approach for the exploration of the giant planets in the Solar System is to focus either on the study of a planetary body and its satellites (e.g. the NASA missions Galileo and Cassini to the Jovian and Saturnian systems) or on the investigation of more specific aspects (e.g. the NASA mission Juno to study the interior of Jupiter and the ESA mission JUICE to explore the Jovian moons Ganymede, Callisto and Europa). This is a well tested approach that allows for a thorough investigation of the subject under study and to collect large quantities of highly detailed data. The only drawback of this approach is that comparative studies of the different giant planets are possible only after decades, especially since the datasets provided by the different missions are not necessarily homogeneous or characterized by the same level of completeness, as the different missions generally focus on different investigations. In the case of the well-studied Jovian and

---

[4] The quoted result has been obtained with a particular choice for the fit. Other, similar estimates have been provided with different fit assumptions.
[5] The only available are those from the Voyager 2 fly-bys.

The Comparative Exploration of the Ice Giant Planets with Twin Spacecraft

Saturnian systems, about 10 years passed before it became possible to compare the dataset supplied by the Galileo mission with the first data supplied by the Cassini mission. Moreover, in order to be able to perform a detailed comparative study of the satellites of these two giant planets it will be necessary to wait until the completion of the JUICE mission, due to the limited coverage of the data from Galileo. As a consequence, about half a century would be required before we can fully address the differences and similarities between the Jovian and Saturnian systems.

Exploring the Uranian and Neptunian systems with the traditional approach would require either

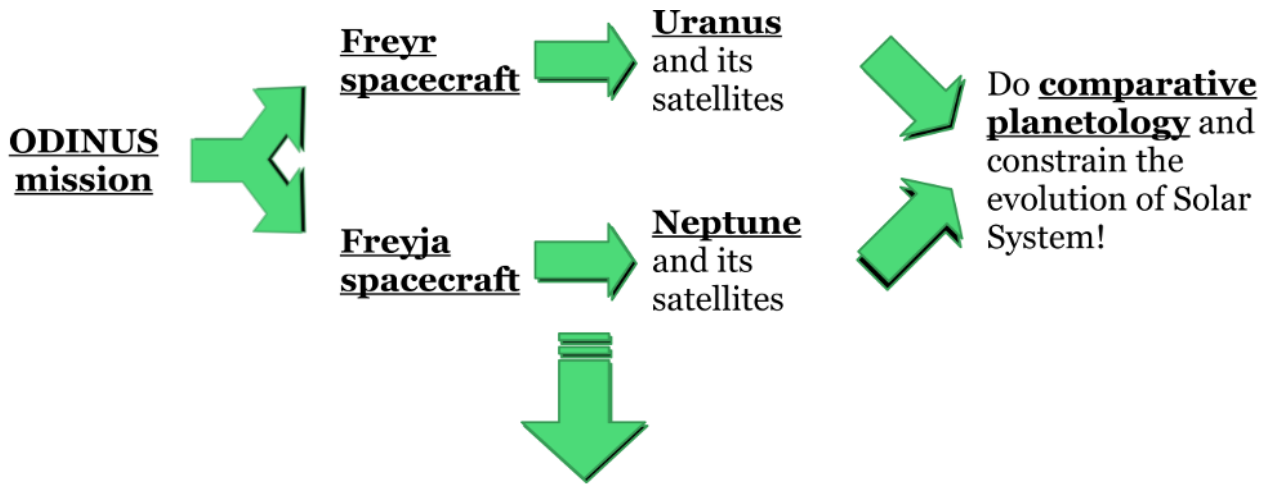

During the cruise(s) ODINUS can perform fundamental physics experiments (gravitation and general relativity in the very-weak field regime) and measurements of the interplanetary medium and solar wind.

*Figure 4: schematics of the twin spacecraft approach of the ODINUS mission.*

half a century of efforts or the focus on this exclusive goal over two consecutive L-class missions of ESA's Cosmic Vision program or its future counterpart. In a scenario where, to balance between the different needs of the astrophysics community as in the recent selection of the science themes for the L2 and L3 missions[6], ESA would devote the L4 and L6 missions to the exploration of these two giant planets, the launch of the L6 missions would occur in 2052 or later (assuming the temporal distance between L4, L5 and L6 is of 6 years as the nominal interval between L2 and L3): assuming travel times to Uranus and Neptune of about 13 and 16 years respectively, as in the scenarios assumed for the Uranus Pathfinder (Arridge et al. 2012) and the OSS (Outer Solar System, Christophe et al. 2012) mission proposals and in the studies conducted by ESA (ESOC 2010) and NASA (Hubbard et al. 2010), the completion of the two missions would occur no earlier than 2068, i.e. more than half a century from now. In the unrealistic scenario of devoting both L4 and L5 missions to the exploration of the ice giants, it would be possible to complete this task by about 2060 but at the cost of not having L-class missions devoted to astrophysics before L6.

The approach proposed by the ODINUS mission concept is different from the traditional one in that it focuses on the use of two M-class spacecraft to be launched toward two different targets in the framework of the same mission (see Fig. 4). The use of twin spacecraft, aside limiting the development cost of the mission, allows for performing measurements with the same set of instruments in the Uranian and Neptunian systems, supplying data of similar quality and potential completeness. Moreover, during at least part of the cruise the two spacecraft will fly on independent orbits, allowing for the study of the interplanetary medium at different angular positions (see Fig. 4). Obviously, the need to produce and manage two spacecraft in place of one limits the amount of instruments that can be included in the scientific payload, implying a less in-depth exploration of the two systems with respect to what would be possible with two dedicated missions. As we discussed in the mission concept that we presented in the white paper "The ODINUS Mission Concept" and that we will now discuss concisely, a careful selection of the instruments and design of the spacecraft can mitigate the importance of this drawback (see Sect. 5.4). Finally, we want to emphasize that, due to the different travel time to reach the two planets, the high-activity phases at the Uranian and Neptunian systems will not overlap (see Sects. 5.1, 5.2 and 5.4), thus limiting the complexity of the mission management.

---
[6] See recommendations of the Senior Survey Committee at http://sci.esa.int/jump.cfm?oid=53261.

## 5.1 The twin spacecraft and their cruise to the ice giants

As we mentioned previously, the founding idea of the ODINUS mission concept is to have a set of twin spacecraft (which we dubbed Freyr and Freyja from the twin gods of the Norse pantheon) to be placed in orbit around Uranus and Neptune respectively (see Fig. 4). In order to fit the budget of an L-class mission, a conservative, straw-man configuration for the ODINUS mission could be based on two spacecraft similar to that of the NASA mission New Horizons, i.e.:

- about 6 instruments in the scientific payload + radio science;
- about 600 kg of dry mass for each spacecraft;
- hybrid (solar electric and chemical) propulsion;
- radioisotope-powered energy source.

The scientific payload and the dry mass of the spacecraft were estimated, in the original white paper, from the assessment of the fuel budget needed to reach the ice giants and to insert them on planetocentric orbits in the worst case scenario. Specifically, we considered the Hohmann transfer orbit between Earth and Uranus (or Neptune) with an orbital insertion at about $2 \times 10^7$ km from the planet on a highly eccentric orbit, obtaining a required $\Delta v$ of about 5 km s$^{-1}$, which in turn translated into a wet-to-dry mass ratio of about five for the spacecraft. This implies that 600 kg of dry mass of each spacecraft requires a wet mass at launch of about 3000 kg. Such a wet mass at launch would make the mission feasible either considering a single launch of the Freyr and Freyja spacecraft with an Ariane V rocket or two separate launches with Soyuz rockets. As we mentioned above, however, this configuration of the spacecraft and of the orbital transfer to the ice giants is extremely conservative and based on the worst case scenario. Preliminary assessments of optimized transfer orbits performed by Thales Alenia Space (J. Poncy, private communication) indicate that a far lower $\Delta v$, estimated to be of the order of 1.5 km s$^{-1}$, could allow for reaching the ice giants and inserting into their orbits while at the same time relaxing the constraints on the wet-to-dry mass ratio.

A related problem is that the studies performed by ESA (ESOC, 2010), NASA (Hubbard et al. 2010) and Thales Alenia Space (J. Poncy, private communication) all indicate that the launch window for a mission to Uranus falls between 2025 and 2030, after which Jupiter will not be in a favourable position. Both studies performed by ESA (ESOC, 2010) and NASA (Hubbard et al., 2010) indicate an even smaller launch window for a mission to Neptune, falling between 2025 and 2028. Launch dates at later times would result in an increase of the time of flight, the required fuel or both. However, these constraints can be loosened with the use of a hybrid solar electric-chemical propulsion system where the solar electric propulsion can be used up to the orbital region of Jupiter in order to by-pass the problem of the unfavourable position of Jupiter and reduce the time of flight (Safa et al. 2013).

A more complete assessment of the orbital path, the wet-to-dry mass ratio and the mass budget for the scientific payload was beyond the scope of the white paper and, more generally, of ESA's call, as the outcome is strongly influenced by the currently undetermined possible launch dates. As a consequence, in this work we maintained the original, conservative estimate of the masses for the

The Comparative Exploration of the Ice Giant Planets with Twin Spacecraft

spacecraft and the scientific payload reported above. In terms of duration of the cruise phases, we adopted nominal times of flight of 13 years for Uranus and 16 years for Neptune based on the results of ESA's (ESOC, 2010) and NASA's (Hubbard et al., 2010) studies. The scenario contemplating two separate launches with Soyuz rockets allows for the two trajectories to be optimized independently, thus allowing for the largest savings of either fuel or travel time. A preliminary check of the orbital positions of Uranus and Neptune showed, however, that the two ice giants will be in a favourable position to launch the two spacecraft together with an Ariane V rocket and then separate their paths between Jupiter and Uranus.

Finally, the distance between the ice giants and the Sun make the use of solar panels difficult in terms of both size and weight of the panels themselves (Arridge et al. 2011; J. Poncy, private communication). A mission to the ice giants therefore requires the use of radioisotope-powered energy source. For a launch date beyond 2034 (the nominal launch date for L3) Am-based thermoelectric generators should be available to European space mission with an adequate technological maturity (TRL 5 foreseen for 2018, Ambrosi 2013). Such generators should be able to provide 1.5-2 W kg$^{-1}$ (Ambrosi 2013; Safa et al. 2013). Assuming a power budget of the order of 200 W for each spacecraft, the mass requested by the power generators and sources would be of the order of 100 kg, thus not affecting in any significant way the mass budget and the possibility to use either Soyuz or Ariane V rockets.

## 5.2 The post-insertion orbital tour of the spacecraft and the exploration strategy of the Uranian and Neptunian systems

In the white paper, the original idea we proposed was to have the spacecraft enter their planetocentric orbits in the regions populated by the irregular satellites thanks to the chemical propulsion, and then to take advantage of ionic propulsion to slowly spiral inward toward the planets. During their inward drift, the spacecraft would have crossed the orbital regions of the regular satellites and, as a end-mission scenario, eventually entered the planetary atmospheres of the two ice giants to perform in situ measurements. However, the energy budget available to the spacecraft, as discussed in Sect. 5.1, would make it impossible to use the ionic propulsion once at Uranus and Neptune (Safa et al. 2013). As a consequence, the orbital tour of the two systems should be realistically planned based only on the use of chemical propulsion and gravitational assists of the satellites, but these constraints should allow for maintaining the basic exploration strategy we proposed in the white paper (Safa et al. 2013).

As mentioned above, the insertion orbits are chosen to insert the spacecraft on high eccentricity orbits at the orbital distance of the irregular satellites, and have one or more fly-bys with members of this family of small bodies. The spacecraft will then change their orbits (either by performing a manoeuvre or taking advantage of a gravitational assist by a regular satellite, thanks to the initial high eccentricity orbit) to transfer to the regions populated by the regular satellites, possibly maintaining eccentric orbits to allow for the contemporary observation of the regular satellites and the planets or their ring systems. The nominal duration of the orbital tours of the two spacecraft, once in orbit around Uranus and Neptune, is planned to be of three years. As three years is also the difference between the duration of the cruise phases of Freyr and Freyja, the Freyr spacecraft would complete its mission at Uranus more or less contemporary to the beginning of Freyja's mission at Neptune. The nominal duration of the orbital tours therefore allows for having only one spacecraft fully operational at a given time, minimizing the complexity of the ODINUS mission in terms of management and optimizing the use of the receiving stations at ground.

In case of a moderate eccentricity of the orbits of the two spacecraft after insertion, the orbital tours at the systems of the ice giants would be divided into two phases: a first 1.5-2 years long phase focusing on the of investigation of the satellites and 1-1.5 years long phase focusing instead

The Comparative Exploration of the Ice Giant Planets with Twin Spacecraft

on the study of the planets and their ring systems. In case, instead, of a high eccentricity of the post-insertion orbits, the two spacecraft could in principle observe the planets and their ring systems while at pericentre and the satellites while farther away from the planets: the orbital tours could then simply be planned as a single 3 years long phase. The high obliquity values of Uranus and Neptune imply that the regular satellites orbit on planes significantly inclined with respect to the ecliptic plane. As a consequence, unless the fuel budget and the orbital studies indicate the possibility of inserting the spacecraft on high-inclination orbits, the orbital paths of the spacecraft will need to be optimized to allow for as many close encounters as possible with the regular satellites in the lifetime of the mission. This is particularly important in the case of Uranus, where the satellites orbit almost perpendicularly to the ecliptic plane: a spacecraft orbiting near the latter would therefore allow only for short close encounters with the regular satellites when they are approaching and crossing the ecliptic plane itself. Based on NASA's studies for a mission to Uranus (Hubbard et al. 2010), a 2 years long phase focused on the exploration of the satellites would allow for two fly-bys of each of the five major satellites of the giant planet.

At the end of the nominal mission at the planets, the updated scenario we propose is to perform a manoeuvre to put the spacecraft on eccentric orbits whose pericentres are located at the boundaries of the atmospheres of the planets, and then perform a second manoeuvre to change their orbits into low eccentricity, high altitude orbits inside the very atmospheres of the planets. The studies performed by NASA for a mission to Uranus (Hubbard et al. 2010) indicate that in the stratosphere of the planet (the same should hold true for Neptune) there is a 300 km wide window where such an atmospheric entry would be feasible without putting at risk the integrity of the spacecraft due to thermal solicitations. This end-mission scenario would allow for performing in situ measurements of the atmospheric compositions and/or densities (depending on the scientific payload, see Sect. 5.3) without putting at risk the other phases of the mission, while at the same time taking advantage of the previous phases of characterization of the planets and their ring systems to minimize the risks damaging the spacecraft due to impacts with dust particles and micrometeorites. Note that the atmospheric entry at Uranus should plausibly occur at one of the poles, at the innermost ring of the planet seems to extend down to the boundary of the planetary atmosphere (De Pater et al. 2013).

## *5.3 The straw-man payload of the twin spacecraft*

A possible straw-man payload for the two spacecraft, which could allow for the achievement of the goals of the ODINUS mission, is composed by:
- Camera (Wide and Narrow Angle);
- VIS-NIR Image Spectrometer;
- Magnetometer;
- Mass Spectrometer (Ions and Neutrals, INMS);
- Doppler Spectro-Imager (for seismic measurements);
- Microwave Radiometer;
- Radio-science package.

The choice to limit the number of instruments on-board the spacecraft is due to the budget constraints, i.e. to the need of keeping the ODINUS mission inside the cost cap of an L-class mission (i.e. about 1 G€). Given the long times required to explore the ice giant planets (i.e. it would take 16 years from launch to explore Uranus and 19 to explore Neptune, see Sects. 5.1 and 5.2), the development of a highly integrated payload would allow for maximizing the number of instruments, thus the scientific return of the mission, and is therefore of critical importance (see also Sect. 5.4). Four instruments that would significantly improve the completeness of the exploration of Uranus and Neptune and their satellites and the scientific return of the mission are:

- Thermal IR Mapper;
- Energetic Neutral Atoms Detector (to complement the measurements of the INMS);
- Plasma Package;
- High-sensitivity Accelerometer (for the post-atmospheric entry phase).

As discussed in Sect. 3.4, the alternative approach based on seismological measurements can be coupled to the more traditional investigation of the gravitational momenta to study the interiors of Uranus and Neptune. Of the two possible approaches (Doppler spectro-imaging or visible photometry) to perform seismological measurements, should visible photometry prove to be the technique of choice, the Doppler Spectro-Imager indicated in the straw-man payload could be replaced by one (or more) alternative instruments. Similarly, a lower wet-to-dry mass ratio than the very conservative one we adopted (see Sect. 5.1) would allow for increasing the dry mass of the spacecraft and, as a consequence, the number of instruments in the scientific payload.

## *5.4 Critical aspects, mitigation strategies and enabling technologies of the ODINUS mission*

The preliminary feasibility assessment of the mission concept performed by ESA's Future Missions Preparation Office evaluated the ODINUS mission as feasible with the budget of an L-class mission for the L3 launch window (Safa et al. 2013) with present-day technology and technologies currently under development in Europe. The two spacecraft are modelled after the one of the ongoing New Horizons mission and their wet masses, according to our first order estimates, would fit either the Soyuz (two launches scenario) or the Ariane V (single launch scenario) payload capabilities. With an estimated final cost of about 550 MEuro (source: NASA[7]) for the New Horizons mission and taking into account that the development costs would be shared between the two spacecraft, the ODINUS mission would be feasible also from the point of view of the expected cost.

The most critical aspects for the success of the ODINUS mission are:
1. the availability of radioisotope-powered energy sources;
2. the achievable transfer rate (mainly in downlink);
3. the achievable wet-to-dry mass ratio and the mass constraints on the scientific payload;
4. the possibility of performing gravitational assist manoeuvres at Jupiter and/or Saturn.

The first two critical aspects is due to the large distances of Uranus and Neptune from the Sun. Concerning the critical aspect 1, said distances make the use of solar panels for energy generation impractical, as already pointed out in Sect 5.1. The development of the required technology and the identification of an affordable and reliable energy source compliant with ESA's policies is therefore mandatory for the feasibility of the ODINUS mission. However, Am-based thermoelectric generators are already under study in Europe and their availability should not present a problem for launch dates later than the nominal one of L2 (Ambrosi 2013; Safa et al. 2013). Concerning critical aspect 2, possible mitigation strategies involve expanding the capabilities of the ESA's network of receiving stations on ground, calibrating the data volume to be collected during the mission phases at the ice giants to the achievable downlink data-rate, or a combination of both. Finally, the critical aspects 3 and 4 are intimately linked but, as we discussed in Sect. 5.1, the adopted wet-to-dry mass ratio is extremely conservative and the use of solar electric propulsion up to the orbital region of Jupiter should allow for by-passing the need for gravitational assist manoeuvres at one of the gaseous giant planets.

---

[7] https://solarsystem.nasa.gov/missions/profile.cfm?MCode=PKB&Display=ReadMore

# 6. ESA's assessment on the scientific theme of the ice giants and conclusions

In the selection for the scientific themes of the L2 and L3 mission, the Senior Survey Committee appointed by ESA stated that "*The SSC considered the study of the icy giants to be a theme of very high science quality and perfectly fitting the criteria for an L-class mission. However, in view of the competition with a range of other high quality science themes, and despite its undoubted quality, on balance and taking account of the wide array of themes, the SSC does not recommend this theme for L2 or L3. In view of its importance, however, the SSC recommends that every effort is made to pursue this theme through other means, such as cooperation on missions led by partner agencies.*" (Cesarsky et al. 2013). With New Horizons well on its path to Pluto and the trans-neptunian region, the ice giants Uranus and Neptune represent indeed the next frontier in the exploration of the Solar System and they potentially hold the key to unlock its ancient past down to its first and more violent phases. Their study can reveal whether the Solar System is one of the possible results of a general scenario of planetary formation, common to all planetary systems, or if the variety of orbital configurations of the extrasolar systems discovered so far are the outcome of a very different sequence of events than those that occurred in the Solar System. In this paper we focused on the scientific rationale of exploring both ice giants in the framework of a single mission, with the goal to perform a comparative study of Uranus, Neptune and their satellite systems. The alternative approach, i.e. the investigation of each of the ice giants with a dedicated space mission, is discussed in the papers by Arridge et al. (this issue) for the case of Uranus, and by Masters et al. (this issue) for the case of Neptune. As we discussed in Sect. 5, these two approaches have different strong and weak points based on the chosen trade-off between depth of exploration and time required to explore both ice giants. Nevertheless, all three mission scenarios designed around these two approaches were deemed feasible, in the framework of the technological development reasonably expected for the L3 launch window, during the preliminary feasibility assessment performed by ESA (Safa et al. 2013). The question we should therefore ask in order to plan the exploration of Uranus and Neptune is not what can we realistically do, but which of the mysteries the ice giants hold the answer to we want to address first.

## Acknowledgements

The authors wish to thank Anna Millillo for valuable discussions and her contribution to improve the measurement strategy, the Italian Space Agency (ASI) for the useful comments and feedback during the preparation of the ODINUS white paper, and the SOC and LOC of the "Uranus beyond Voyager 2" workshop for the useful discussions during the meeting. The authors also wish to thank Davide Bellotto, Maurizio Pajola, Mark Hofstadter, Heidi Hammel and all the other supporters of the ODINUS white paper for their enthusiasm and encouragement across the preparation of the white paper and the whole selection process of the scientific themes of the L2 and L3 missions by ESA. Finally, the authors would like express their gratitude to Mauro Turrini, Vanda Zottarelli and Donatella Festa for creating the public outreach material for the ODINUS white paper, and to Mirko Riazzoli and Danae Polychroni for their advices and comments.

## Bibliography

- Ambrosi R., 2013. "European Space Nuclear Power Systems: Enabing Technology for Space Exploration Missions". Abstract presented at the "Uranus beyond Voyager 2" Workshop held in Meudon, France, on 16-18 September 2013.
- Anderson, J.D. et al. 1989, Astrophys. J. 342, 539-544

The Comparative Exploration of the Ice Giant Planets with Twin Spacecraft

- Anderson, J.D. et al. 1995, Astrophys. J. 448, 885-892
- Anderson, J.D. et al. 1998a, Icar. 131, 167-170
- Anderson, J.D. et al. 1998b, Phys. Rev. Lett. 81, 2858-2861
- Arridge C. et al. 2012, Experimental Astronomy 33, 753-791
- Axford, W.I., 1973, Space Sci. Rev. 14, 582
- Baines, K.H. et al. 1995, Icarus, vol. 114, no. 2, p. 328-340
- Ballester, G.E., in: Wamsteker, W., Gonzalez Riestra, R. (eds.) Ultraviolet Astrophysics Beyond the IUE Final Archive, Proceedings of the Conference held in Sevilla, Spain, from 11–14 November 1997, ESA SP, vol. 413, p. 21. ESA Publications Division (1998)
- Barr, A. C., Canup, R. M. 2010, Nature Geoscience 3, 164-167.
- Belcher, J. W., et al. 1989, Science, 246, 1478,
- Bertotti, B et al. 2003, Nature 425, 374-376
- Bishop, J., S.K. Atreya, P.N. Romani, G.S. Orton, B.R. Sandel, and R.V. Yelle, The middle and upper atmosphere of Neptune, in Neptune and Triton, edited by D.P. Cruikshank, p.427, Univ. Arizona Press, Tucson (1995).
- Borucki W.J., Koch D., Jenkins J., et al., 2009, Science, 325, 709
- Broadfoot, A.L., SK. Atreya, J.L. Bertaux, J.E. Blamont, A.J. Dessler, et al., Ultraviolet spectrometer observations of Neptune and Triton, Science, 246, 1459 (1989).
- Buratti, B.J., Mosher, J.A., 1991. Icarus 90, 1–13.
- Capone, L.A., Whitten, R.C., Prasad, S.S., Dubach, J. 1977, Astrophys. J. 215, 977–983
- Cesarsky C., Benz W., Bertolucci S., Bignami G., Encrenaz T., Genzel R., Spyromilio J., Zarnecki J., 2013, "Report of the Senior Survey Committee on the selection of the science themes for the L2 and L3 launch opportunities in the Cosmic Vision programme", accessible on ESA's website at http://sci.esa.int/ssc_report.
- Chandler, M.O., Waite, J.H. 1986, Geophys. Res. Lett. 13, 6–9
- Christophe B. et al. 2012, Experimental Astronomy 34, 203-242
- Connerney, J. E. P., Acuna M. H. and Ness N. F. (1991) The magnetic field of Neptune. J. Geophys. Res., 96, 19023-42.
- Coradini, A., Magni, G., Turrini, D. 2010, Space Science Reviews 153, 411-429.
- Coradini, A., Turrini, D., Federico, C., Magni, G. 2011, Space Science Reviews 163, 25-40.
- Dalton, J. B., Cruikshank D. P., Stephan K., McCord T. B., Coustenis A., Carlson R. W., Coradini A., 2010, Space Science Reviews 153, 113-154.
- Damour, T., Piazza, F., Veneziano, G., 2002, Phys. Rev. Lett. 89, 081601.
- De Pater, I., Lissauer J.J. 2010. Planetary Science - Second Edition. Cambridge University Press, Cambridge, UK
- De Pater I., et al., 2013. "Uranus main ring system". Abstract presented at the "Uranus beyond Voyager 2" Workshop held in Meudon, France, on 16-18 September 2013.
- Decker, R.B., Cheng, A.F. 1994, JGR, 99, E9, 19027–19045
- European Space Operations Centre (ESOC), 2010, "Interplanetary transfers to the outer planets with probe release for the timeframe 2025-2035", accessible on ESA's website at http://sci.esa.int/jump.cfm?oid=47682.
- Famaey, B., & McGaugh, S.S. 2012, Living Rev. Relativity, 15, (2012), 10. [Online Article]: cited [2013-12-12], http://www.livingreviews.org/lrr-2012-10
- Feuchtgruber, H. et al. 1994, Nature, 389, pp. 159-162

- Feuchtgruber, H. et al. 2013, Astronomy & Astrophysics, Volume 551, id.A126, 9 pp.,
- Fichtner, H. et al., 2000. AIP Conf. Proc. 528, 345
- Fitzpatrick, P.J. et al., 2014, Astrophysics and Space Science, 350, pp 65-88.
- Fressin F., Torres G., Charbonneau D., and the Kepler Team, 2013, presentation at the 221st American Astronomical Society Meeting held on 6-10 January 2013 in Long Beach, CA.
- Folkner, W.M., Williams, J.G., Boggs, D.H. 2008, JPL Memorandum 343R-08-003
- Funsten, H.O. et al., 2009. Science 326, 964
- Gaulme P., Mosser B., 2005, Icarus, 178, 84
- Gaulme P., Schmider F.X., Gay J., Guillot T., Jacob C., 2011, A&A, 531, A104
- Goldreich, P., & Keeley, D. A. 1977, ApJ, 212, 243
- Gomes R., Levison H.F., Tsiganis K., Morbidelli A., 2005, Nature, 435,466-469.
- Gómez-Pérez, N., Heimpel, M., 2007. Numerical models of zonal flow dynamos: An application to the ice giants. Geophys. Astrophys. Fluid Dyn. 101, 371–388.
- Grundy W. M., Young L. A., Spencer J. R., Johnson R. E., Young E. F., Buie M. W., 2006, Icarus, 184, 543-555.
- Guervilly, C., Cardin, P., Schaeffer, N., 2012. A dynamo driven by zonal jets at the upper surface: Applications to giant planets, Icarus 218, 100-114.
- Gruntman, M.A. et al., 2001. J. Geophys. Res. 106, 15767
- Gudkova, T.V. et al. 1988, Astronomicheskii Vestnik, vol. 22, p. 23-40
- Guillot, T., 2005, Ann. Rev. Earth Planet. Sci. 33, 493–530
- Hartmann, W. K., Ryder, G., Dones, L., Grinspoon, D. 2000, in Origin of the Earth and Moon 493-512.
- Hammel, H.B. et al., 2001, Icarus, 153, 229–235.
- Hammel, H.B. et al., 2005, Icarus, 175, 534–545.
- Helled, R., Anderson, J.D., Podolak, M., Schubert, G. 2011, Astrophys. J. 726, 15, 7
- Herbert, F., J. Geophys. Res. 114, A11206 (2009)
- Holme, R., Bloxham, J., 1996. The magnetic fields of Uranus and Neptune: Methods and models. J. Geophys. Res. 101, 2177–2200.
- Hsieh, K.C., Shih, K.L., Jokipii, J.R., Grzedzielski, S., 1992. Astrophys. J. 393, 756
- Hubbard, W.B., Nellis, W.J., Mitchell, A.C., Holmes, N.C., McCandless, P.C., Limaye, S.S., 1991. Interior structure of Neptune – Comparison with Uranus. Science 253, 648–651.
- Hubbard, W.B., Podolak, M., Stevenson, D.J., 1995. The interior of Neptune. In: Cruikshank, D.P., Matthews, M.S., Schumann, A.M. (Eds.), Neptune and Triton. The University of Arizona Press, Tucson, AZ, pp. 109–138.
- Hubbard W. B., and the Ice Giants Decadal Study Team, 2010, "Ice Giants Decadal Study", accessible on the website of the National Academies at http://sites.nationalacademies.org/SSB/SSB_059331.
- Iafolla, V. et al., Plan. Space Sci. 58, 300-308 (2010)
- Jewitt, D., Haghighipour, N. 2007, Ann. Rev. of Astron. and Astrophys., 45, 261-295.
- Kaspi, Y., W. B. Hubbard, A. P. Showman, and G. R. Flierl (2010), Gravitational signature of Jupiter's internal dynamics, Geophys. Res. Lett., 37, L01204, doi:10.1029/2009GL041385
- Krimigis, S. M., et al., Science, 246, 1483, 1989
- Krimigis et al., 2009. Science, 326, 971-973

The Comparative Exploration of the Ice Giant Planets with Twin Spacecraft

- Krivov A. V., Wardinski I., Spahn F., Kruger H., Grun E., 2002. Icarus 157, 436
- Lellouch E., et al., A&A 518, id. L152, 4 pp., 2010.
- Levison, H. F., Bottke, W. F., Gounelle, M., Morbidelli, A., Nesvorny, D., Tsiganis, K. 2009, Nature 460, 364-366.
- Levison, H. F., Morbidelli, A., Tsiganis, K., Nesvorny, D., Gomes, R. 2011, AJ 142, 152.
- Lissauer, J.J., and Stevenson, D.J., 2007, in Protostars and Planets V, Reipurth, Jewitt, and Keil, (Eds.). Univ. of Arizona Press.
- Lunine J. I., 1993, Ann. Rev. of Astron. and Astrophys., 31, 217-263.
- Madhusudhan N., Seager S., ApJ 729, article id. 41, 13 pp., 2011.
- Mauk, B. H., E. P. Keath, M. Kane, S. M. Krimigis, A. F. Cheng, M. H. Acufia, T. P. Armstrong, and N. F. Ness, J. Geophys. Res., 96, 19,061, 1991
- Mauk, B. H., S. M. Krimigis, A. F. Cheng, and R. S. Selesnick, in Neptune and Triton, edited by D. Cruikshank, University of Arizona Press, Tucson, in press, 1994
- Majeed, T., Waite, J.H., Bougher, S.W., Yelle, R.V., Gladstone, G.R., McConnell, J.C., Bhard-waj, A., Adv. Space Res. 33(2), 197–211 (2004)
- Martin, S. C. et al., Astrophysics and Space Science, 337 (1), 2011
- Marzari F., Weidenschilling S. J. 2002, Icarus 156, 570-579
- Matter, A., Guillot, T., Morbidelli, A. 2009. Planetary and Space Science 57, 816-821
- McComas, D.J. et al., 2009. Space Sci. Rev. 146, 11
- Michel E., Baglin A., Auvergne M., et al., Oct. 2008, Science, 322, 558
- Möbius, E., et al., 2009. Direct observations of interstellar H, He, and O by the Interstellar Boundary Explorer, Science, 326(5955), 969–971, doi:10.1126/science.1180971.
- Morbidelli A., Levison H. F., Tsiganis K., Gomes R., 2005, Nature 435, 462-465.
- Morbidelli, A., Tsiganis, K., Crida, A., Levison, H. F., Gomes, R. 2007, AJ 134, 1790-1798.
- Mosqueira, I., Estrada, P., Turrini, D. 2010. Space Science Reviews 153, 431-446.
- Murphy N., et al., "ECHOES – Sounding Jupiter's Atmosphere & Interior" instrument study submitted to NASA in response to AO NNH12ZDA006O.
- Nellis, W.J., Holmes, N.C., Mitchell, A.C., Hamilton, D.C., Nicol, M., 1997. Equation of state and electrical conductivity of "synthetic Uranus," a mixture of water, ammonia, and isopropanol, at shock pressure up to 200 GPa (2 Mbar). J. Chem. Phys. 107, 9096–9100.
- Ness, N. F., M. H. Acuna, K. W. Behannon, L. F. Burlaga, J. E. P. Connerney, R. P. Lepping, and F. M. Neubauer, Magnetic fields at Uranus, Science, 233, 85-89, 1986.
- Ness, N., Acuna, M.H., Burlaga, L.F., Connerney, J.E.P., Lepping, R.P., Neubauer, F.N., 1989. Magnetic Fields at Neptune, Science, Vol. 246 no. 4936 pp. 1473-1478.
- Nesvorny, D. 2011. Young Solar System's Fifth Giant Planet?. ApJ 742, L22.
- Nesvorny, D., Vokrouhlicky, D., Morbidelli, A. 2007. AJ 133, 1962-1976.
- Owen, T., Mahaffy, P., Niemann, H. B., Atreya, S., Donahue, T., Bar-Nun, A., de Pater, I. 1999, Nature 402, 269-270.
- Papaloizou, J. C. B., Nelson, R. P., Kley, W., Masset, F. S., Artymowicz, P. 2007, in Protostars and Planets V, ed. B. Reipurth, D. Jewitt, & K. Keil, Univ. Arizona Press, Tucson, AZ, 655
- Pearl, J.C. et al., Icarus, vol. 84, 1990, p. 12-28
- Pearl, J.C. et al., Journal of Geophysical Research Supplement, vol. 96, Oct. 30, 1991, p. 18,921-18,930

The Comparative Exploration of the Ice Giant Planets with Twin Spacecraft

- Podolak, M. et al., Planetary and Space Science v. 43, p. 1517-1522, 1995
- Prockter, L. M., et al. 2010. Space Science Reviews 153, 63-111.
- Redmer, R., Mattsson, T.R., Nettelmann, N., French, M., 2011. The phase diagram of water and the magnetic fields of Uranus and Neptune. Icarus 211, 798–803.
- Richardson, J. D., J. W. Belcher, M. Zhang, and R. L. McNutt, J. Geophys. Res., 96, 18,993, 1991
- Safa F., and the ESA Future Missions Preparation Office, 2013. Preliminary feasibility assessment of the ODINUS mission concept.
- Safronov, V. S. (1969). Evolution of the protoplanetary cloud and formation of the earth and planets. Translated from Russian in 1972. Keter Publishing House, 212 pp.
- Schmider F.X., Gay J., Gaulme P., et al., Nov. 2007, A&A, 474, 1073
- Schmider F. X. et al. (2009). Study proposal for the instrument DSI – DoppleR Spectro-Imager submitted to the Call for Payload Study Consortia for Cosmic Vision L-class Mission EJSM/Laplace.
- Schubert, G., Hussmann, H., Lainey, V., Matson, D. L., McKinnon, W. B., Sohl, F., Sotin, C., Tobie, G., Turrini, D., van Hoolst, T. 2010. Space Science Reviews 153, 447-484.
- Selesnick, R.S., Richardson, J.D., Geophys. Res. Lett. 13, 624–627 (1986)
- Martin et al., 2012, Astrophys. Space Sci. 337, 65-78
- Sittler Jr., E.C., Ogilvie, K.W., Selesnick, R., J. Geophys. Res. 92, 15263 (1987)
- Soderlund K. M., Heimpel M. H., King E. M., Aurnou J. M., Icarus 224, pp. 97-113, 2013.
- Spencer J. R., et al., in Saturn from Cassini-Huygens, Dougherty M. K., Esposito L. W., Krimigis S. M., Springer, London (UK), 2009, pp 683-724 .
- Sromovsky L.A., et al., Icarus 105, 110–141, 1993.
- Sromovsky, L.A. et al., Icarus 220, pp. 694-712, 2012.
- Stanley, S., Bloxham, J., 2006. Numerical dynamo models of Uranus' and Neptune'smagnetic fields. Icarus 184, 556–572.
- Sumi T., et al. Nature 473, 349-352, 2011.
- Suomi, V.E. et al., Science, vol. 251, Feb. 22, 1991, p. 929-932
- Tamayo D., Burns J. A., Hamilton D. P., Hedman M. M. 2011. Icarus 215, 260-278
- Tamayo D., Burns J. A., Hamilton D. P. 2013. Icarus 226, 655-662.
- Trafton, L.M., Miller, S., Geballe, T.R., Tennyson, J., Ballester, G.E. 1999, ApJ 524, 1059–1083
- Tosi F., Turrini D., Coradini A., Filacchione G., and the VIMS Team 2010. MNRAS 403, 1113-1130.
- Tsiganis K., Gomes R., Morbidelli A., Levison H. F., 2005, Nature, 435,459-461.
- Turrini D., Magni G., Coradini A. 2011, MNRAS 413, 2439-2466.
- Turrini D., Coradini A., Magni G. 2012, ApJ 750, 8.
- Turrini D. 2013. Planetary & Space Science. DOI: 10.1016/j.pss.2013.09.006i.
- Turrini D., Nelson R. P., Barbieri M., 2014, Experimental Astronomy, DOI: 10.1007/s10686-014-9401-6.
- Turrini D., Svetsov V. 2014, Life 4, 4-34, DOI: [10.3390/life4010004](10.3390/life4010004).
- Turyshev, S.G., Annu. Rev. Nucl. Part. Sci. 58, 207-248 (2008)
- Vasyliunas, V.M., Geophys. Res. Lett. 13, 621–623 (1986)
- Verbiscer A. J., Skrutskie M. F., Hamilton D. P., 2009. Nature 461, 1098.

The Comparative Exploration of the Ice Giant Planets with Twin Spacecraft

- Walsh K. J., Morbidelli A., Raymond S. N., O'Brien D. P., Mandell A. M., 2012, M&PS, 47, 1941
- Walsh K. J., Morbidelli A., Raymond S. N., O'Brien D. P., Mandell A. M., 2011, Nature, 475, 206
- Weidenschilling, S. J. 1975, Icarus 26, 361-366
- Weidenschilling, S. J., Davis, D. R., Marzari, F. 2001. Earth, Planets, and Space 53, 1093-1097
- Weidenschilling S. J., Marzari F., 1996, Nature, 384, 619-621
- Will, C.M. 2006, Living Rev. Relativity, 9, (2006), 3. [Online Article]: cited [2013-12-12], http://www.livingreviews.org/lrr-2006-3
- Wong M. H., Lunine J. I., Atreya S. K., Johnson T., Mahaffy P. R., Owen T. C., Encrenaz T., 2008. Reviews in Mineralogy & Geochemistry, 68, 219-246, DOI: 10.2138/rmg.2008.68.10.
- Zarka, P., B.M. Pederson, A. Lecacheux, M.L. Kaiser, M.D. Desch, W.M. Farrell, and W.S. Kurth, Radio emissions from Neptune, in Neptune and Triton, edited by D.P. Cruikshank, p.341, Univ. Arizona Press, Tucson (1995).